\documentclass{edp-jp4}
\usepackage[T1]{fontenc}
\def\eg{{\it e.g. }}
\def\etal{{et al. }}  
\def\ie{{\it i.e. }}

\def\zu{\rm\,}     

\def\Hcm{\hbox{$\,{\rm H\, cm^{-2}}\,$}}

\begin{document}
\title{Archeops: A CMB anisotropy balloon experiment  \\ measuring
        a broad range of angular scales}
\author{F.-Xavier Désert}
\address{Laboratoire d'Astrophysique, Observatoire de Grenoble, BP 53,\\ 
414 rue de la piscine, 38041 Grenoble Cedex 9, France\\
Francois-Xavier.Desert@obs.ujf-grenoble.fr, www.archeops.org}
\author{the Archeops Collaboration}\address{from the following institutes:\\
California Institute of Technology, Pasadena USA\\
Centre d'Etude Spatiale des Rayonnements, Toulouse France\\
Centre de spectrométrie nucléaire et de spectrométrie de masse, Orsay France\\
Colllège de France, Paris France\\
DAPNIA CEA, Saclay France\\
Institut d'Astrophysique de Paris, Paris France\\
Institut d'Astrophysique Spatiale, Orsay France\\
Institut des Sciences Nucléaires de Grenoble, Grenoble France\\
IROE CNR, Firenze Italy\\
Jet Propulsion Laboratory, Pasadena USA\\
Laboratoire de l'accélérateur linéaire, Orsay France\\
Landau Institute of Theoretical Physics, Moscow Russia\\
Observatoire Midi-Pyrénées, Toulouse France\\
Queen Mary and Westfield College, London UK\\
Universita di Roma La Sapienza, Roma Italy\\
University of Minnesota at Minneapolis USA\\
}
%
\maketitle
\begin{abstract}
  The Cosmic Microwave Background Radiation is the oldest photon
  radiation that can be observed, having been emitted when the
  Universe was about 300,000 year old. It is a blackbody at 2.73~K,
  and is almost perfectly isotropic, the anisotropies being about one
  part to 100,000. However, these anisotropies, detected by the COBE
  satellite in 1992, constrain the cosmological parameters such as the
  curvature of the Universe.
  
  Archeops is a balloon-borne experiment designed to map these
  anisotropies. The instrument is composed of a 1.5~m telescope and
  bolometers cooled at 85~mK to detect radiation between 150 and
  550~GHz. To lower parasitic signals, the instrument is
  borne by a stratospheric balloon during the arctic night.  This
  instrument is also a preparation for the Planck satellite mission,
  as its design is similar to HFI.
  
  We discuss here the results of the first scientific flight from
  Esrange (near Kiruna, Sweden) to Russia on January 29th 2001, which
  led to a 22\% (sub)millimetre sky coverage unprecedented at this
  resolution. Here, we put some emphasis on interstellar dust foreground
  emission observations.
\end{abstract}

\section{The scientific objective}

The Cosmic Microwave Background Radiation (CMBR) shows some small
temperature differences of the order of one part in 100000, that were
measured for the first time by the COBE satellite \cite{Smoot}. These
so-called anisotropies trace the fluctuations of the density of matter
that are thought to be the origin, by gravitational collapse, of the
large-scale structure of the Universe (galaxies, clusters,...) that we
observe today. Its pattern can also yield an indirect measurement of
the density, age and curvature of the Universe (see \eg\cite{Hu}).
There have been many experiments that have already measured these
anisotropies with various techniques, angular resolution, noise and
scanning strategy. Mosqt recent ones (e.g. TOCO,
Boomerang~\cite{deBernardis,Netterfield}, and
Maxima~\cite{Hanany,Lee}) have improved on COBE results by the
wavelength coverage, the sensitivity and the angular resolution.

The Archeops experiment aims at mapping the anisotropies of the cosmic
microwave background from small to large scales at the same time. For
this purpose, a beam of about 8 arcminutes is swept through the sky by
spinning a 1.5~m telescope pointing at 41~degree elevation around its
vertical axis. A large fraction of the sky is covered when the
rotation of the Earth makes the swept circle drift across the
celestial sphere. This is only possible if the observations are done
during the Arctic night and on a balloon where neither the Sun nor the
atmosphere disturb the measurements. Ozone cloud emission and residual
winds can be avoided with a high altitude strastospheric balloon. From
the Swedish balloon and rocket base in Esrange near Kiruna, in
cooperation with Russian scientists, the CNES balloon team can launch
balloons in the polar night, with a typical trajectory ending just
before the Ural mountains in Russia. Integration times can be up to
24~hours in the December-January campaigns.

\section{The instrument}

A general description of the first Archeops instrument can be found in
\cite{Benoit} where the first gondola used during the test flight
(that happened in Trapani in July 1999) is described. The present
experiment uses the same concept, details of which are given in \cite{Bnt2}.

\subsection{The telescope, optics and detectors}

The Archeops telescope is a two mirror, off-axis, tilted Gregorian
telescope consisting of a parabolic primary (1.5~m main diameter) and
an elliptical secondary. The telescope was designed to provide
diffraction-limited performance when coupled to single mode horns
producing beams with FWHM of 8~arcminutes or less at frequencies
higher than 140~GHz.

For CMB anisotropy measurements, control of spectral leaks and beam
sidelobe response is critical. Archeops channels have been
specifically designed to maximize the sensitivity to the desired
signal, while rejecting out-of-band or out-of-beam radiation. We have
chosen to use the configuration developed for Planck HFI, using a
triple horn configuration for each photometric pixel. In this scheme,
radiation from the telescope is focussed into the entrance of a
back-to-back horn pair (10~K stage). It creates a beam-waist where
wavelength selective filters can be placed (1.6~K stage). Finally, the
third horn (0.1~K stage) maintains beam control and focuses the
radiation onto the bolometer placed at the exit aperture.
A convenient aspect of this arrangement is that the various components
can be placed on different temperature stages in order to create
thermal breaks and to reduce the level of background power falling
onto the bolometer and fridge.

Twenty two spider-web bolometers are placed on the 100~mK low
temperature plate.  There are 9 bolometers at 143~GHz, 7 at 217~GHz ,
6 polarised bolometers at 353~GHz and two at 545~GHz, placed at
different points in the focal plane. They observe the same sky pixel
at different times, from 100~msec to a few minutes. The six 353~GHz
channels are devoted to the measurement of galactic polarized
emission. These are assembled in three pairs, with one single
back-to-back horn and a polarizer splitter (the so-called OTM
configuration) for each pair. The two bolometers of each pair measure
the polarized intensity of the incoming signal in two orthogonal
directions. Each pair makes a different polarising angle with respect
to the scan axis to enable the full determination of the Stokes
parameters. Archeops will provide the first measurement of
polarization in this range of frequencies with a sensitivity adequate
for measuring galactic dust polarised emission, a CMB foreground as
yet unmeasured for the preparation of Planck-HFI.

\subsection{The gondola, the pivot and the fast star sensor}
\label{sec:gondola}

The gondola is made with welded aluminium square tubes and a careful
design prevents important deformations of the optical design in the
presence of strength.  The two mirrors and the cryostat are fixed to
the frame. The pivot connects the flight chain of the balloon to the
payload through a thrust bearing, providing the necessary degree of
freedom for payload spin. It includes a torque motor that acts
against the flight chain to spin the payload.  The rotation of the
payload is controlled via a vibrating structure rate gyroscopes that
can detect angular speeds as low as 0.1~deg/s.

A custom star sensor has been developed for pointing reconstruction in
order to be fast enough to work on a payload rotating at 2-3~rpm.  We
have developed a simple night sensor, based on a telescope with
photodiodes along the boresight of the mm-wave telescope. Thus, like
the millimeter telescope, the star sensor scans the sky along a circle
at an elevation of 41~deg. A linear array of 46 sensitive photodiodes
were placed in the focal plane of a 40~cm diameter, 1.8~m focal length
parabolic optical mirror. The line of photodiodes is perpendicular to
the scan and covers 1.4~degrees in elevation on the sky. We can
observe during one rotation of the payload stars up to magnitude 7 \ie
between 50 and 100 stars per turn during night time. An optical filter
allows this star sensor to yield at least a few detected stars even in
the presence of low elevation Sun. Pointing reconstruction is done a
posteriori by comparing star candidates and a dedicated star catalog.
The precision of the pointing solution is better than 1~arcminute rms
for the Trapani test flight and 2~arcmin. for the Kiruna 2001 flight.

\subsection{The cryogenics, the electronics and the telemetry}
\label{sec:cryo}

The focal plane is cooled to 100~mK by means of an open cycle dilution
refrigerator. This type of refrigerator has been designed for
satellite applications (it will be used on Planck HFI) and Archeops is
the first balloon-borne experiment using a dilution refrigerator.  The
bolometers are placed on the 100~mK stage supported by Kevlar cords.
The bolometers are biased using AC square waves by a capacitive
current source. Their output is measured with a differential
preamplifier (the first stage uses JFET working at about 120~K) and
digitized before demodulation, with boxes already designed in
preparation for Planck HFI. Data ampled at 152~Hz are
compressed and stored in a 2~Gbyte on-board flash-eprom memory. A
telemetry channel using the Inmarsat satellite is used to control the
experiment during the whole flight.

\section{Archeops flights and first results}
\label{sec:flight}

A first flight of the instrument took place in Trapani on July 17th
1999. This test flight used only a few detectors (5) and we got only 4
hours of data during the night. Nevertheless, this flight allowed us
to check all the fonctionnalities of the instrument~\cite{Benoit}.
Preliminary results concerning the Galactic Plane emission are shown by Boulanger
\etal (this conference).

One flight was successful at the end of the Dec~2000-Jan~2001 campaign
in Kiruna (Sweden), on the 29th January 2001; it lasted 7h30 at a
32~km altitude. Very high stratospheric winds limited both the flight
duration and the altitude. The (sub)millimetre beams could be measured
during the flight when the telescope crossed Jupiter twice and are as
expected (optical beam of 8~arcmin. at 143, 217 and 545~GHz and
6~arcmin. at 353~GHz).  The 143 and 217~GHz signals are dominated by
the cosmic dipole and the 10~K back-to-back horn emission (sinusoidal
shape). At 353 and 545~GHz, the emission from the Galaxy is dominant
as well as some atmospheric signal. A significant fraction of the sky
could be observed (22\%) albeit with a small zone covered twice. The
galactic plane is well observed at all frequencies from the
anticenter to the Cygnus regions. Some clouds much below the Galactic
plane can easily be identified with their CO and infrared counterparts
(Perseus, Taurus, Pleiades).  In-flight calibrations with the CMB
dipole and the Galaxy as measured by COBE-FIRAS agree within 10\% of
each other. Sensitivity to cirrus HI clouds is estimated at 545~GHz as
$2\times 10^{20}\Hcm (\theta/1\zu deg)^{-1}$ ($1\ \sigma$) for square
areas with an angular side of $\theta$ and standard dust emissivities.

With the 353~GHz channels, Archeops will provide the first measurement
of galactic polarized emission in this type of frequencies. It is an
important topic in the prospect of foreground removal for Planck-HFI,
and is also of great interest to constrain the physics of galactic
dust and molecular clouds. The sensitivity for the current flight is a
degree of 5\% polarisation ($1\ \sigma$) for $A_V=15$ in a one-square
degree patch.  Sensitivities are typically between 50 and $100\zu \mu
K_{RJ}$ for one second of integration and for one photometric pixel at
143 and 217~GHz. There are about 8 pixels with a CMB sensitivity
between 120 and $200 \zu \mu K_{CMB}$ for one second of integration
and for one photometric pixel.

Good detections of the CMB anisotropy spectrum can be expected from
large angular scales to beyond the first so-called acoustic peak. This
work is currently in progress. Archeops should also be able to constrain
dust emissivity laws in the many galactic regions that were not
resolved by FIRAS (a hotly debated issue in this conference). Another
specific benefit from Archeops is to connect, in the spherical
harmonic $l$-space sense, the calibration of the FIRAS low resolution
all-sky (sub)millimetre survey up to the high resolution small area
observations from ground telescopes.

The development of this current Archeops project owes a lot to the
pionneering work and enthusiasm of Guy Serra in the domain of
submillimetre astronomy.  We also wish to thank the CNES and Esrange
Swedish Facility for their continued support for this project and the
flights (technical and scientific) that were realised very smoothly.


\begin{thebibliography}{9}
\bibitem{Benoit}
Benoît, A., \etal, Astroparticle Physics, in press, astro-ph/0106152, 2001

\bibitem{Bnt2}
Benoît, A., and the Archeops collaboration, 2001, ESA-SP-471, Proc. of
\emph{the 15th ESA Symposium on European Rocket and Balloon Programmes and
Related Research}, p. 431

\bibitem{deBernardis}
de Bernardis, P., \etal, Nature, 404, 955, 2000


\bibitem{Hanany}
Hanany, S., \etal,  Astrophysical Journal, 343, L3, 2000

\bibitem{Hu} 
Hu, W., Sugiyama, N., and Silk, J., Nature, 386, 37, 1997

\bibitem{Lee}
Lee, A. T., \etal, preprint, astro-ph/0104459, 2001

\bibitem{Netterfield} Netterfield, C. B., \etal, Astrophysical
  Journal, submitted, astro-ph/0104460, 2001
  

\bibitem{Smoot}
Smoot, G. F., \etal, Astrophysical Journal, 371, L1, 1991

\end{thebibliography}
\end{document}